# Narrowband Internet of Things for Non-terrestrial Networks

Olof Liberg, Stefan Eriksson Löwenmark, Sebastian Euler, Björn Hofström, Talha Khan, Xingqin Lin, and Jonas Sedin

Ericsson

{olof.liberg, stefan.g.eriksson, sebastian.euler, bjorn.hofstrom, talha.khan, xingqin.lin, jonas.sedin}@ericsson.com

*Abstract*—The Narrowband Internet of Things (NB-IoT) is a cellular access technology developed by the Third Generation Partnership Project (3GPP) to provide wide area connectivity for the Internet of Things. Since its introduction in 3GPP Release 13, NB-IoT has in a few years achieved a remarkable market presence and is currently providing coverage in close to 100 countries. To further extend the reach of NB-IoT and to connect the unconnected, 3GPP Release 17 will study the feasibility of adapting NB-IoT to support non-terrestrial networks (NTNs). In this article, we review the fundamentals of NB-IoT and NTN and explain how NB-IoT can be adapted to support satellite communication through a minimal set of modifications.

## Introduction

In the Third Generation Partnership Project (3GPP) Release 15, the Fifth Generation (5G) cellular system was specified. It is designed to support an unprecedented versatility. While earlier cellular systems have focused on providing voice and mobile broadband (MBB) services in public networks, 5G strives to broaden the range of supported use cases. 3GPP Release 15 addresses massive Internet of Things (IoT) deployments using Narrowband Internet of Things (NB-IoT) and Long-Term Evolution (LTE) for Machine Type Communication (LTE-M), and critical IoT applications using 5G New Radio (NR) ultra reliable and low latency communication (URLLC) [1]. Release 16 strengthens the support for industrial verticals by specifying operation in license-exempt frequency bands and support for non-public networks.

Release 17 will further extend the 5G outreach by including solutions to enable NR operation in non-terrestrial networks (NTNs) using airborne or spaceborne platforms [2]. NR will be adapted to support communication using low earth orbit (LEO) and geosynchronous orbit (GEO) satellite constellations. It is anticipated that the adaptions will provide the tools to also support high-altitude platform station (HAPS) and air-to-ground (A2G) communication. The focus of Release 17 NTN work is to enable MBB and fixed wireless access in areas where it is commercially challenging to provide terrestrial network coverage. It builds on previous study items including a Release 16 study on identifying efficient solutions for making NR ready for satellite communication [3].

Release 17 will also study the feasibility of adapting NB-IoT and LTE-M to support NTN [4]. NB-IoT NTN will seemingly face many of the challenges identified for NR NTN [3]. The details of the devised solutions may differ since NR and NB-IoT use different radio interfaces. Although this work may reuse parts of the Release 16 NTN study performed for NR, we emphasize that NB-IoT and LTE-M by design are systems of low complexity compared to NR. This has influenced the scope of the NB-IoT and LTE-M NTN study item and is also expected to affect its conclusions. To make NB-IoT NTN and LTE-M NTN attractive technologies for the cellular industry, it is important to limit the scope of the needed adaptations.

In this article, we present our views on this study with a focus on NB-IoT and describe how NB-IoT can be modified to support both LEO and GEO based NTNs through a set of essential adaptations.

## NB-IoT

NB-IoT is envisaged to serve massive IoT applications and devices of low complexity. It is optimized for services characterized by small, delay-tolerant and infrequent data transmissions. It is designed to meet 5G massive IoT performance requirements in terms of:

- Enhanced coverage for supporting devices deployed in deep indoor environments such as basements.
- Power-efficient operation for facilitating a beyond 10-year device battery life.
- Support for massive number of devices making small and infrequent data transmissions.

NB-IoT is an LTE-based technology and can simply be described as a narrowband version of LTE. It supports a 180 kHz carrier bandwidth which matches that of an LTE physical resource block. It supports two types of carriers: anchor and non-anchor. The anchor carrier supports cell-defining broadcast transmissions including synchronization and system information signaling which configures the radio access network. Non-anchor carriers can be activated for improving the system access and data transmission capacity beyond that offered by the anchor carrier.

The NB-IoT physical layer uses orthogonal frequency-division multiplexing (OFDM) in the downlink (DL), while the uplink (UL) is based on discrete Fourier transform-precoded OFDM to limit the peak-to-average power ratio of UL transmission. It uses a DL transmission bandwidth of 180 kHz and can tolerate a DL signal-to-noise ratio (SNR) down to -14.5 dB [5]. The UL transmission bandwidth ranges between 3.75 and 180 kHz. This relatively low bandwidth was selected to optimize the UL SNR for facilitating efficient operation in challenging coverage conditions. NB-IoT supports an UL SNR of -13.8 dB for a transmission bandwidth of 15 kHz [5].

The NB-IoT higher layers are based on the LTE protocol stack, and are divided into a user plane (UP) and a control plane (CP). The UP manages data transmissions and is defined by the Packet Data Convergence Protocol (PDCP), Radio Link Control (RLC) and Medium Access Control (MAC) protocols. The CP is responsible for signaling and supports the Radio Resource Control (RRC) protocol in addition to PDCP, RLC and MAC. In NB-IoT, the CP functionality has been extended to support small and infrequent data transmissions.

NB-IoT supports a coupling loss, which is the signal attenuation between transmitting and receiving nodes' antenna connectors, of beyond 164 dB. This corresponds to a coverage extension of around 20 dB compared to earlier 3GPP technologies. This improvement is supported by a flexible transmission time interval (TTI) for all the physical channels. In the extreme case, transmissions of up to 20 seconds are possible to support cell edge connectivity.

NB-IoT can be deployed to operate within an LTE carrier, in the guard band of an LTE carrier, or in dedicated spectrum since Release 13. Since Release 15, NR can also cater to NB-IoT inband operation. NB-IoT is even capable of connecting to the 5G Core (5GC) network since Release 16. Thus, NB-IoT qualifies as a 5G technology as it can operate in the 5G spectrum, connect to the 5GC and meet the 5G massive IoT performance requirements [5].

## NTN

Satellite communication is (re)gaining traction both in the cellular industry and academia. This is fueled by bold plans to launch thousands of satellites in LEO constellations as advertised, for example, by OneWeb [6], SpaceX [7] and Kuiper [8]. Although the integration of NB-IoT into NTN may appear as new in the context of 3GPP, both academia [9] [10] and industry [11]-[13] have already studied this topic. In [9], a time-frequency Aloha scheme was investigated for supporting unsynchronized unidirectional communication between NB-IoT devices and a LEO satellite under a low system load. In [10], the Doppler frequency shift observed in LEO constellations was studied, and a resource allocation scheme was proposed to reduce its impact on the NB-IoT radio interface.

In 3GPP, studies on satellite network uses cases [11], architectural aspects [12], radio channel models [13], and solutions for adapting NR to support NTN have been completed [3]. These studies were performed in the context of 5G with the 5GC and 5G NR in mind. As a large portion of the work is generic, it forms a solid foundation for the Release 17 study on NB-IoT and LTE-M for NTN. The overarching objective in this study is to provide truly ubiquitous IoT connectivity, focusing on locations where terrestrial network coverage is absent. NB-IoT and LTE-M for NTN should thus be seen as a complement to existing terrestrial deployments, and not as a competitor.

Various architectural options for integrating a 3GPP radio into a satellite system exist. Figure 1 illustrates a satellite operating according to the bent-pipe architecture, which is the focus of the 3GPP NB-IoT NTN study. The figure shows that the NB-IoT base station (BS) is integrated in the gateway. A possible option for connecting BS and gateway is to use Common Public Radio Interface (CPRI). The gateway feeds the satellite with a modulated signal over the feeder link. The satellite payload is said to be transparent, which means that it acts as a sophisticated repeater that filters, amplifies, and shifts the signal from the feeder link frequency band to the service link frequency band. The gateway transceiver is stationary, powerful and typically dimensioned to make the feeder link non-limiting. Consequently, the focus of this article is on the service link performance. The figure illustrates a single spotbeam, that is the coverage area provided by a beam, which is commonly associated with a cell. A single satellite may in practice support several beams. The figure also presents the elevation angle of the satellite link. This is an important parameter that influences the slant range, the line of sight probability, and the satellite velocity relative to the device.

Another option is to use regenerative architecture where the satellite payload contains the cellular BS. It may offer advantages such as reduced latency, implementation flexibility and improved signal quality. Compared to the regenerative architecture, however, the bent-pipe architecture can shorten the time to market by expediting the integration of NB-IoT in NTN. Already deployed satellite constellations supporting a transparent payload may be upgradable to support NB-IoT. In contrast, new satellite constellations need to be deployed for the regenerative case.

Satellite constellations come in many flavors. 3GPP has focused on GEO and LEO constellations, with the remark that HAPS requirements constitute a subset of the requirements defined by GEO and LEO systems. In other words, HAPS will be supported by any 3GPP radio that supports LEO and GEO satellite communication.

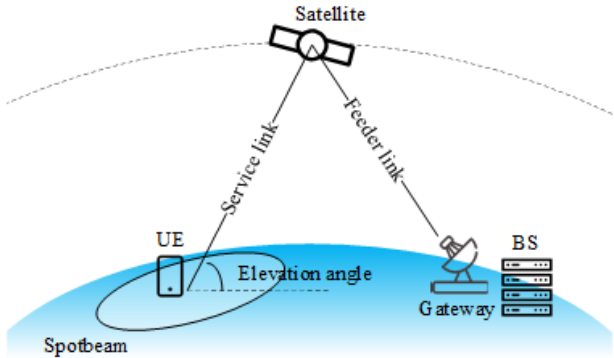

Figure 1 Satellite with transparent payload operating according to the bent-pipe architecture.

*Geosynchronous orbit*

GEO is characterized by an orbital period of 24 hours, synchronous with the earth's rotation. A special GEO case is the geostationary orbit, where the satellite is placed in the equatorial plane at an orbit altitude of 35,786 km. In the eyes of an observer on earth, a geostationary satellite appears to be operating from a fixed position in the sky. Each GEO satellite is assigned an orbital slot, which corresponds to a distinct longitudinal position at the mentioned altitude. To avoid interference between adjacent GEO satellites communicating with earth stations on shared frequency bands, angular separations of the orbital slots are enforced.

Most GEO satellites are located in the equatorial plane, that is in geostationary orbits. However, to counteract inhomogeneities of the earth's gravitational field, tidal forces by the Sun and the Moon, and other effects, a satellite needs to perform station-keeping using thruster burns to secure its longitudinal and latitudinal position. To save thruster fuel, which is a scarce resource that may limit a satellite's operational life, some satellites only perform east-west station keeping. This will lead to a gradual inclination of the satellite orbit relative to the equatorial plane. Due to the inclination, the satellite will no longer be stationary relative to an observer on earth. The projection of the satellite trajectory on earth recorded over a day will form a figure-8 pattern. In astronomy, this phenomenon is known as an analemma.

Based on the 3GPP technical reports [3][13] and some key characteristics of a geostationary NTN, assuming bent-pipe architecture, operation in the S-band at a carrier frequency of 2 GHz, and an elevation angle spanning between 10° and 90°, the following parameters values can be determined:

- Spotbeam diameter of 100 – 3500 km.
- Round-trip time (RTT) from the gateway to the device, and back in the range of 477 – 541 ms.

The expected *SNR* for a signal bandwidth *BW* can be determined as a function of the transmitter equivalent isotropically radiated power (*EIRP*), the receiver antenna gain-to-noise-temperature (*G/T*), and the signal attenuation which in turn depends on the free space path loss (*FSPL*), scintillation loss (*SL),* atmospheric loss (*AL*), shadow fading (*SF*) and Boltzmann's constant *k* [3], using

$$SNR = EIRP + G/T - 10\log_{10}(k) - FSPL - SF - SL - AL - 10\log_{10}(BW). \quad (1)$$

In Table 1, based on the parameters considered in [3], we estimate the expected NB-IoT SNR for GEO NTN to be between 0 and 1.27 dB in the DL, and -8 to -6.7 dB in the UL.

Table 1 NB-IoT link budgets for GEO and LEO NTNs assuming an S-band carrier frequency of 2 GHz.

|  | GEO | | LEO | |
| --- | --- | --- | --- | --- |
|  | DL | UL | DL | UL |
| EIRP | 51.6 dBW | -7 dBW | 26.6 dBW | -7 dBW |
| G/T | -31.6 dB/K | 19 dB/K | -31.6 dB/K | 1.1 dB/K |
| BW | 180 kHz | | 180 kHz | |
| FSPL | 189.5 – 190.6 dB | | 154 – 164.2 dB | |
| SL | 2.2 dB | | 2.2 dB | |
| AL | 0.03 – 0.2 dB | | 0.03 – 0.2 dB | |
| SF | 3 dB | | 3 dB | |
| SNR (dB) | 0.04 – 1.27 | -7.96 – -6.73 | 1.44 – 11.8 | 0.54 – 10.9 |

To support general geosynchronous orbits, we next turn the attention to the Doppler shift observed for a satellite operating in an inclined orbit. Figure 2 shows how an inclination results in a time-of-day dependent Doppler shift that repeats in a diurnal cycle. The results depend on the inclination angle and the device latitude, as they affect the elevation angle and the speed of the satellite relative to the observer. In this example, we assume a 10º inclination and an observer at latitude 59° North, which corresponds to the location of Stockholm, Sweden. A Doppler shift ranging up to 500 Hz is observed in this example which assumes operation in the S-band, at 2 GHz carrier frequency.

In short, the main challenges for adapting NB-IoT to a GEO NTN stem from the high RTT and pathloss, the large cell size, and the Doppler shift associated with an inclined orbit. Later, we will discuss these aspects in the context of NB-IoT and present their impact on the NB-IoT system design.

*Low earth orbit*

LEO satellites are located at an altitude ranging between 500 and 2000 km, with orbital periods between 94 and 127 minutes. We follow 3GPP's example and focus on the LEO constellations at 600 km altitude, for which a satellite moves at a velocity of 7.56 km per second relative to earth.

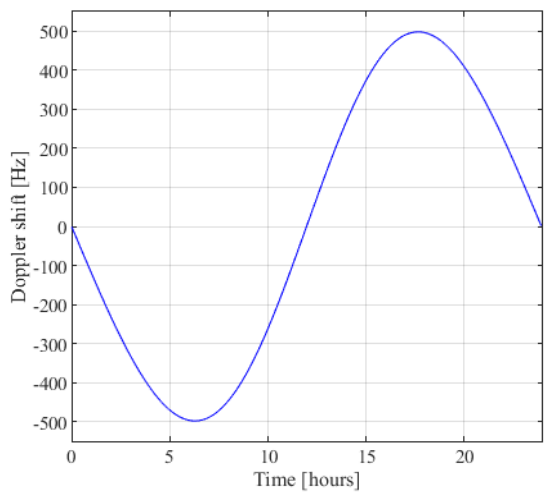

Figure 2 Doppler frequency shift observed at latitude 59° North for a GEO satellite operating at an inclination angle of 10°.

The high velocity of the satellite induces a significant Doppler shift on the service link. To limit the maximum Doppler shift, a lower bound is typically defined for the satellite's elevation angle. In 3GPP, it is assumed that the service link is operational for elevation angles exceeding 10°. This agreement limits the maximum Doppler shift to 24 ppm for the case of 600 km LEO. The Doppler shift is not constant as it changes with the elevation angle. It can be shown that the Doppler shift changes approximately linearly within a spotbeam. Figure 3 shows the linear Doppler frequency variation within an earth-moving spotbeam of diameter 50 km generated by a LEO satellite operating at 600 km altitude using a carrier frequency of 2 GHz.

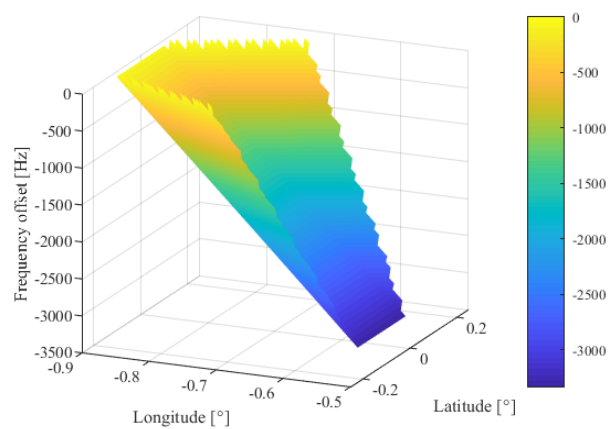

Figure 3 DL Doppler frequency variation within a moving spotbeam generated by a LEO satellite at 600 km in the S-band.

There are two types of deployments envisioned for LEO: earth-moving beams and earth-fixed beams. In the former case, the satellite beams are constantly moving along with the satellite, and a satellite spotbeam can be expected to provide coverage for a fixed point on earth for just a few seconds. In the latter case, the satellite adapts its beamforming mechanism to realize spotbeams that remain fixed on earth. It can be shown that a satellite at 600 km altitude can support an elevation angle, defined at the cell center, exceeding 10° for around 450 seconds [14]. To guarantee continuous coverage, a service link switch will need to be performed very frequently for the moving beam scenario and once every few minutes for the fixed beam case.

The RTT for a LEO at 600 km altitude ranges between 8 and 25.8 ms for the bent-pipe architecture. The satellite motion causes the satellite-to-device delay to vary at a rate proportional to the Doppler shift, that is up to 24 µs/s.

The spotbeam diameter for LEO 600 km can be assumed to range between 50 and 1000 km [3]. This is lower than that for GEO, but still significantly larger than what is anticipated for a terrestrial network. The SNR is also more favorable for LEO than GEO due to the lower *FSPL* which, assuming an S-band carrier frequency of 2 GHz, equals 154 dB and 164.2 dB for elevation angles of 90° and 10°, respectively. As shown in Table 1, the expected NB-IoT SNR range for LEO NTN is around 1.4 to 11.8 dB for DL, and 0.5 to 10.9 dB for UL.

For LEO NTN, the main challenge for NB-IoT is to manage aspects related to the high satellite velocity including the Doppler shift, its variation and the variation of the timing. In Table 2, based on the parameter values considered in [3], we summarize the key characteristics of GEO and LEO NTN in the context of NB-IoT.

Table 2 An overview of LEO (600 km) and GEO NB-IoT NTN assuming bent-pipe architecture.

|  | **GEO** | **LEO** |
|---|---|---|
| Altitude | 35786 km | 600 km |
| Spotbeam radius | 100 – 3500 km | 50 – 1000 km |
| Minimum elevation angle | 10° | 10° |
| RTT | 477 – 541 ms | 8 – 25.8 ms |
| Maximum differential delay within a beam | 3.12 ms | 10.3 ms |
| Maximum Doppler shift | 24 ppm | 0.93 ppm |
| Example data rate | 100s of bps to 10s of kbps | 1 – 100 kbps |
| Example applications | Utility metering, environmental monitoring, remote sensing. | Wearables, fleet management, tracking. |

## NB-IoT SUPPORT FOR NTN

NB-IoT has in just a few years achieved an impressive commercial presence with an ecosystem consisting of core and radio network suppliers, device vendors, and mobile network operators. NB-IoT networks currently provide coverage in close to 100 countries [15]. For NB-IoT and NTN to be an attractive combination in the existing ecosystem, it is important to limit the set of NB-IoT modifications needed to support NTN. The NB-IoT design in Release 13 through 16 has followed a principle of low complexity, which should also be honored when evolving the technology to support NTN. In this section, we describe the main challenges facing NB-IoT NTN and discuss potential solutions.

*Architectural aspects*

The Release 17 work on NB-IoT NTN will focus on the bent-pipe architecture as it may facilitate integration of NB-IoT into existing NTNs based on satellites with transparent payload. From a 3GPP perspective, the transparent payload satellite can be viewed as a sophisticated repeater. Unlike a basic repeater, it can also support frequency shifting from the feeder link to the service link frequency band. The 3GPP specifications need to be evolved to support this type of repeater node for NB-IoT.

LEO constellations support earth-moving or earth-fixed beams. Since NB-IoT does not support RRC connected mode mobility, the frequent beam switching expected in the moving beam case will force a device to move back and forth between RRC idle and RRC connected states. To avoid this behavior, 3GPP is expected to prioritize the earth-fixed beam architecture.

NB-IoT is supported by the Evolved Packet Core (EPC) since Release 13, and by the 5GC since Release 16. As existing deployments are supported by EPC, it is natural to prioritize EPC for NB-IoT over NTN. It is expected that EPC should be able to support NB-IoT for NTN without any significant changes needed for the core functionality.

3GPP has agreed to assume that NTN NB-IoT devices will be equipped with a Global Navigation Satellite System (GNSS) receiver. Earlier, GNSS support for NB-IoT had not been anticipated since the GNSS chip increases the overall modem cost which is critical in terrestrial massive IoT applications. Given that the current NTN device price level by far exceeds that expected in terrestrial networks, GNSS support appears to be a reasonable assumption for NTN NB-IoT. A device may use its GNSS location information in combination with satellite ephemeris data to support mobility, compensate for Doppler effects, and achieve time and frequency synchronization. Basic satellite ephemeris information can be stored in the device's universal subscriber identity module (uSIM) and complemented by more detailed information broadcasted in the NTN.

*Physical layer aspects*

NB-IoT supports LTE frequency division duplex (FDD) and time division duplex (TDD) frequency bands in the range of 450 to 2690 MHz. To limit the impact on the current design, it is beneficial if the support for NTN is limited to the existing frequency range supported by LTE which ends at roughly 6 GHz. NB-IoT FDD bands should be prioritized since TDD is challenging to support in NTNs with large RTTs.

Timing is an important aspect in NTNs that needs careful consideration. For both GEO and LEO satellites, the spotbeam radius as well as the device-BS link distances can be significantly larger than the cell radius of 120 km typically assumed in terrestrial NB-IoT. This will impact mechanisms for time/frequency adjustment including Timing Advance (TA) and UL frequency compensation indication. GNSS-aided solutions can be devised to address these issues. For instance, the NTN RTT by far exceeds what NB-IoT is designed for. As a GNSS-equipped device can estimate and pre-compensate the RTT before random access, the BS receiver need only deal with a small residual timing error. We next discuss these aspects in detail.

A device may use its GNSS-acquired geographical location in combination with satellite ephemeris data to estimate the service link propagation delay. By compensating for the estimated delay at the transmission of the random access preamble, UL time synchronization is achieved already from the first transmission. The initial UL TA command sent from the network in the Random Access Response (RAR) message can be modified to support a bipolar timing range to allow for correction of GNSS measurement inaccuracy. The RAR reception timing relative to the preamble transmission timing needs to be adjusted to account for the maximum supported RTT in the NTN.

The device sends, for example, the RRC Connection Request message to acknowledge the RAR reception. The RAR transmission to RRC Connection Request reception scheduling gap also needs to be dimensioned according to the maximum supported RTT. The RRC Connection Request message can be updated to include the propagation delay estimated by the device, which the network may use for device-tailored scheduling of UL and DL transmissions. The dynamic scheduling control information needs to be updated to account for the large RTT. A simple alternative is to extend the control to data channel scheduling delay with an offset corresponding to the NTN RTT.

To compensate for the satellite mobility, a device may be required to autonomously perform TA updates in RRC connected mode based on the satellite ephemeris information. This is particularly relevant in LEO networks where the propagation delay can change at a considerable rate.

Location information provided by the satellite ephemeris may also support prediction of Doppler shift and Doppler shift variation rate. A device can use this information to compensate for the expected Doppler shift at DL signal reception and at UL transmission. NB-IoT supports a TTI of more than 4 seconds in the UL and 20 seconds in the DL. For long TTIs, the device will need to continuously update the frequency compensation due to Doppler shift and the timing for reception and transmission. NB-IoT is designed for low mobility and negligible Doppler shift, so GNSS-assisted Doppler compensation will be useful for supporting both LEO and GEO satellites in inclined orbits.

The earlier discussion showed that the expected UL SNR for GEO systems will be as low as -8 dB for a transmission bandwidth of 180 kHz. NB-IoT does by design support operation in power-limited conditions, for example by means of bandwidth adaptation for improved transmit power spectral density. For instance, the UL SNR can be improved by reducing the transmission bandwidth to 15 kHz. This will improve the UL SNR for GEO deployments by around 10 dB compared to that presented in Table 1. This indicates that it may be possible to support low data rate GEO satellite communication using the 23 dBm device power class. If 180 kHz UL transmissions are to be supported efficiently, that is using short TTIs, then the device can be equipped with a directional antenna providing a gain on the order of 10 dBi.

NTN can experience high levels of interference when operating at high traffic loads. This is due to a relatively slow gain roll-off of the cell-defining beam lobes, which leads to inter-beam and inter-cell interference [3]. For interference mitigation, it is common to configure different carrier frequencies in adjacent cells. Currently, NB-IoT device measurement capabilities are limited to three measurement frequencies which in turn limits the frequency reuse factor to 1/3. This number should preferably be increased to ensure that NB-IoT supports efficient RRC idle mode mobility in an NTN with frequency reuse.

*Higher layer aspects*

For many NB-IoT devices, power-efficient operation in RRC idle mode is important. Since a device is expected to determine its position to estimate the service link propagation delay before initiating the random access procedure, it is natural to also use this information for accurate cell selection and reselection. A cell is suitable for selection only if the estimated RTT is below the maximum RTT assumed by the network in the RAR timing configuration. Furthermore, a device can rank all detected cells based on its geographical distance to the corresponding cell centers. Selecting to camp on the geographically closest cell will guarantee that a device camps on the best cell from an SNR perspective.

Most procedures in a cellular network are controlled by various timers defined in the MAC, RLC, PDCP and RRC layers. For NB-IoT, these timers are already dimensioned quite generously to support operation in extended coverage which requires long TTIs. The impacts due to the high RTT are therefore limited. We have only identified three timers that need to be optimized for NTN:

- The MAC Contention Resolution timer which starts at the RRC Connection Request message transmission and defines the window during which the device should monitor for the RRC Connection Setup message. It takes a maximum value of 10.24 s.

- The MAC hybrid automatic repeat request (HARQ) RTT timer which, for example, is started after UL data transmission is completed and whose expiration triggers monitoring of the DL control channel for scheduling of a retransmission.

- The RLC t-reordering timer which is started in case RLC packets are received out of order. Its expiration triggers the generation of an RLC status report.

The start of the first two timers can be delayed by the RTT for reducing the device monitoring time and improving the device power efficiency. The t-reordering timer has a maximum value of 1600 ms which can be extended to better support the high RTTs expected in a GEO NTN.

NB-IoT supports at most two HARQ processes. A HARQ process requires that a first transmission is acknowledged before a second transmission is scheduled on the same HARQ process. In an NTN where the minimum time between initial transmission and received acknowledgment is determined by the RTT, this implies that only two data transmissions, one for each HARQ process, can be made every RTT. Increasing the number of HARQ processes is not a viable approach as it will negatively impact the devices' receiver complexity. Thus, an important enhancement is to allow HARQ to be disabled and let RLC automatic repeat request (ARQ) manage the retransmissions of failed transmissions. The RLC layer supports transmission of multiple blocks before feedback needs to be sent, and therefore allows for a significantly higher data rate than what can be supported with HARQ. Further, for a transmission with HARQ disabled, a lower target block error rate can be used for robustness.

Finally, to support the functionalities outlined in this article, the cell-defining RRC layer broadcast signaling needs to be updated with NTN-specific information including:

- Satellite ephemeris data defining the locations of the satellites in a constellation.

- The maximum RTT assumed for the RAR and RRC Connection Request timing configuration.

- The geographical center point of each cell used for cell selection and reselection.

Table 3 provides a summary of the solutions discussed in this section.

Table 3 Overview of key issues facing NB-IoT NTN and potential solutions.

| Issues | Solutions |
|---|---|
| Initial synchronization amid large RTT and Doppler shift | Use GNSS and ephemeris data to estimate and pre-compensate for RTT and Doppler shift before random access |
| High delay and Doppler variation rate | Device performs autonomous TA and frequency adjustments |
| Low UL SNR | Use narrow bandwidth or directional antennas |
| High interference | Enhance inter-frequency measurement capability of NB-IoT devices |
| Higher layer timers | Offset Contention Resolution and HARQ RTT times by an RTT and extend RLC t-ordering timer |
| Limited number of HARQ processes | Disable HARQ and rely on RLC retransmissions |
| Cell (re)selection | RTT based cell suitability determination, and distance-based cell ranking |
| NTN-specific signaling support | Use RRC broadcast signaling for updating ephemeris, maximum supported RTT, etc. |
| Transparent payload | Evolve specification to support advanced repeater node with frequency conversion capability |

## CONCLUSIONS

In this article, we have briefly introduced NB-IoT and described how its design can be enhanced to support satellite communication. We have identified a limited set of adaptations to the physical layer functionality that addresses random access and RRC connected mode transmission timing, Doppler compensation, and enhanced inter-frequency measurements. For the higher layer design, we have suggested to base the RRC idle mode mobility on satellite ephemeris and device positioning information, adjust the range or delay the start of three vital timers, and optionally disable HARQ. Finally, we have proposed NTN-specific RRC signaling to support the new functionality. With these adaptations, we expect that NB-IoT will support both GEO and LEO earth-fixed beam constellations. The bent-pipe architecture is most attractive as it admits the existing NB-IoT technology to the NTN evolution while expediting commercialization.

NB-IoT and LTE-M are complementary technologies that can address different types of IoT use cases based on their unique capabilities. While we have focused on NB-IoT NTN in this article, a natural extension is to study the feasibility of LTE-M NTN and identify how LTE-M can be modified to support NTN with minimal impact on the existing design.

## BIOGRAPHIES

OLOF LIBERG (olof.liberg@ericsson.com) holds a bachelor's degree in Business and Economics and a master's degree in Engineering Physics, both from Uppsala University. He joined Ericsson in 2008 and has specialized in the standardization of cellular radio access technologies. He was the chairman of 3GPP TSG GERAN during the 3GPP study on new radio access technologies for Internet of Things leading up to the specification of NB-IoT, and is currently leading Ericsson's 3GPP radio access network standardization team. He is the leading author of the book Cellular Internet of Things (Elsevier), and has contributed to several academic articles and US patents.

STEFAN ERIKSSON LÖWENMARK (stefan.g.eriksson@ericsson.com) holds a master's degree in Computer Science and Electrical Engineering from Linköping University, Sweden. He joined Ericsson in 1997 and has contributed as a researcher, research leader and standardization delegate in numerous areas including GSM/EDGE, EC-GSM-IoT and 5G NR. He has contributed to academic articles and several US patents. He is currently working as a 3GPP RAN1 delegate in the area of non-terrestrial 5G networks.

SEBASTIAN EULER (sebastian.euler@ericsson.com) is a Senior Researcher at Ericsson Research, Sweden. He joined Ericsson in 2016 and has since focused on extending the LTE and 5G New Radio (NR) standards with support for satellite networks and aerial vehicles. He has contributed to academic articles and US patents. He has a background in particle physics, and received his Ph.D. from RWTH Aachen University, Germany, in 2014. Before joining Ericsson, he held a postdoctoral position at Uppsala University, during which time he worked with neutrino experiments in Antarctica.

BJÖRN HOFSTRÖM (bjorn.hofstrom@ericsson.com) holds a master´s degree in electrical engineering from Linköping University, Sweden, including one year at the University of Massachusetts Amherst, US. He has worked as a researcher and standardization delegate where he contributed to the work on 5G NR, EC-GSM-IoT and GSM/EDGE technologies. He is a currently working as a Research Leader at Ericsson and has contributed to academic articles and several US patents.

TALHA KHAN (talha.khan@ericsson.com) received his M.S.E. and Ph.D. degrees in electrical and computer engineering from The University of Texas at Austin, USA, and his B.Sc. degree in electrical engineering from the University of Engineering and Technology Lahore, Pakistan. He is a currently working as a Senior Researcher at Ericsson Research Silicon Valley, USA. Before joining Ericsson, he has held summer internship positions at Broadcom, Mitsubishi Electric Research Labs and Connectivity Lab, Facebook. His research interests include cellular systems, non-terrestrial networks, stochastic geometry applications and energy harvesting.

XINGQIN LIN (xingqin.lin@ericsson.com) is a Master Researcher and Standardization Delegate at Ericsson, leading research and standardization in the areas of non-terrestrial networks including satellite, high-altitude platform station, aircraft, and drones. He has contributed to 5G NR, NB-IoT, and LTE standards. He is co-author of the book "Wireless Communications and Networking for Unmanned Aerial Vehicles." He served as an editor of the IEEE COMMUNICATIONS LETTERS from 2015-2018. He holds a Ph.D. in electrical and computer engineering from The University of Texas at Austin, USA. He is an IEEE Senior Member.

JONAS SEDIN (jonas.sedin@ericsson.com) holds a master's degree in electrical engineering from KTH (Royal Institute of Engineering), Sweden. He joined Ericsson Research in 2018 focusing on evaluating and extending standard support for 5G NR in aerial and satellite networks, as well as acting as a delegate in 3GPP RAN2. He is currently a member of IEEE 802.11 working with the next generation of Wi-Fi networks.